\documentstyle[aps,epsfig]{revtex}

\bibstyle{unsrt}
\begin{document}

\title{Derivative Expansion and Soliton Masses}
\author{Gerald V. Dunne}
\address{Department of Physics, University of Connecticut, Storrs CT
06269, USA}

\maketitle
\vskip .5cm

\begin{abstract} We present a simple algorithm to implement the
generalized derivative expansion introduced previously by L-H. Chan, and
apply it to the calculation of the one-loop mass correction to the
classical soliton mass in the 1+1 dimensional  Jacobi model. We then show
how this derivative expansion approach implies that the total (bosonic
plus fermionic) mass correction in an N=1 supersymmetric soliton model is
determined solely by the asymptotic values (and derivatives) of the
fermionic background potential. For a static soliton the total mass correction is $-m/(2\pi)$, in agreement with recent analyses using phase-shift methods. 
\end{abstract}

\vskip 1cm

The calculation of quantum corrections to classical soliton masses is a
key ingredient in the semiclassical approach to quantum field theory
\cite{rajaraman,rebbi}. This question has been re-addressed recently
\cite{top,jaffe} for $1+1$ dimensions using topological boundary
conditions and phase-shift methods. In this Letter I compute the quantum
mass correction for solitons in the recently introduced Jacobi model
\cite{dunne}, using a simple algorithm based on the derivative expansion
\cite{aitchison,chan}; and I apply this derivative expansion approach to
the total mass correction in an $N=1$ SUSY model in $1+1$ dimensions.

Consider the following Lagrangian for a real scalar field $\phi$ in 1+1
dimensions:
\begin{eqnarray} {\cal L}=\frac{1}{2}(\partial_\mu \phi)(\partial^\mu\phi) 
- V(\phi)
\label{blag}
\end{eqnarray} 
We shall consider the examples:
\begin{eqnarray}
V(\phi)=\cases{\frac{m^4}{8\lambda}\left[\left(\frac{\sqrt{\lambda}}{m}
\phi\right)^2  -1\right]^2\qquad   ; \phi^4 {\rm model}\cr
\frac{m^4}{\lambda}\left[1-\cos\left(\frac{\sqrt{\lambda}}{m}\phi\right) 
\right]\qquad   ; {\rm SineGordon\,\, model}\cr
\frac{2 m^4}{\lambda}\, {\rm sn}^2\left(\frac{\sqrt{\lambda}}{2m}\phi| \nu\right) \qquad   ;  {\rm Jacobi \,\,model}}
\label{pots}
\end{eqnarray} 
The $\phi^4$ and Sine-Gordon models are text-book cases
\cite{rajaraman}, while  the Jacobi potential has recently been studied in the context of 
an exactly solvable model of quantum mechanical instantons \cite{dunne}.
Since quantum mechanical instantons have the same functional form as
solitons in 1+1 dimensions, some of the results of \cite{dunne} carry
over here. The function ${\rm sn}(z|\nu)$ is one of the Jacobi elliptic
functions \cite{ww}, and $0\leq\nu\leq 1$ is the (real) elliptic parameter. This parameter $\nu$ controls the shape and period of the Jacobi potential, as illustrated in Figure 1. The Jacobi model is a deformation of the Sine-Gordon model, reducing smoothly to Sine-Gordon as $\nu\to 0$ [note that 
${\rm sn}(\frac{\phi}{2}|\nu=0)=\sin(\frac{\phi}{2})$]. 

As is well known \cite{rajaraman}, the mass $m$ and coupling $\lambda$ may
be scaled out, so that the semiclassical loop expansion is an expansion
in powers of $\frac{\hbar\lambda}{m^2}$. With this understood, we can set
$m=\lambda=1$ in our one-loop calculations.

These models have classical static soliton/antisoliton solutions 
$\phi_c(x)$ satisfying
\begin{eqnarray}
\phi_c^\prime(x)=\pm\sqrt{2 V(\phi_c(x))}
\label{soliton}
\end{eqnarray} 
For the potentials in (\ref{pots}) the classical solitons are
\begin{eqnarray}
\phi_c(x)=\cases{{\rm tanh}(\frac{x}{2})\qquad   ; \phi^4 {\rm model}\cr
4\, {\rm arctan}(e^x)\qquad   ; {\rm SineGordon\,\, model}\cr 
2K(\nu)+2\,{\rm sn}^{-1}({\rm tanh}(x))\qquad   ;  {\rm Jacobi \,\,model}}
\label{sols}
\end{eqnarray} 
Here $K(\nu)=\int_0^{\pi/2}d\theta/\sqrt{1-\nu\sin^2\theta}$, is the elliptic
quarter period \cite{ww}. Note that $K(0)=\pi/2$, and
$K(\nu)\sim\frac{1}{2}\log(\frac{16}{1-\nu})$ as $\nu\to 1$. 

\begin{figure}[h]
\vskip -1cm
\centering{\epsfig{file=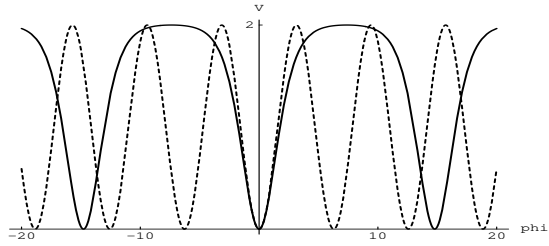,width=5in,height=4.5in}}
\vskip -2.5in
\caption{The Jacobi potential $V(\phi)=2\,{\rm sn}^2(\frac{\phi}{2}|\nu)$. The solid curve is for $\nu=0.99$, while the dashed curve is for $\nu=0$, which is just the Sine-Gordon case.}
\end{figure}

The classical mass of the soliton is given, in units of 
$\frac{m^3}{\lambda}$, by
\begin{eqnarray} 
M_{\rm classical}=\int_{\rm min}d\phi\sqrt{2 V(\phi)}
=\cases{\frac{2}{3}\qquad   ;\phi^4 {\rm model}\cr  
8 \qquad   ; {\rm SineGordon\,\, model}\cr
\frac{4}{\sqrt{\nu}}\log\left(\frac{1+\sqrt{\nu}}{1-\sqrt{\nu}}\right)
\qquad   ;  {\rm Jacobi \,\,model}}
\label{cmasses}
\end{eqnarray} 
where the integration is between the two neighboring minima of $V(\phi)$  between which the soliton interpolates.

\begin{figure}[h]
\vskip -1cm
\centering{\epsfig{file=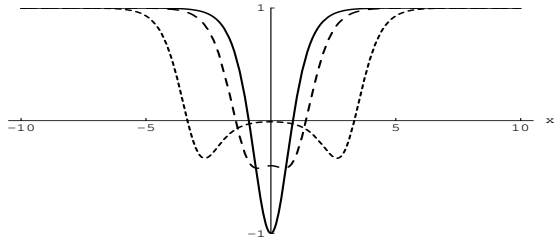,width=5in,height=4.5in}}
\vskip -2.5in
\caption{The bosonic fluctuation potential $V^{\prime\prime}(\phi_c(x))$
for the Jacobi model. The solid curve is for $\nu=0$, which is the same
as the Sine-Gordon case. The dashed curve is for $\nu=0.6$, and the
dotted curve is for $\nu=0.99$. When $\nu\neq 0$, this fluctuation
potential is not of the solvable Po\"schl-Teller form.}
\end{figure}

In units of $m$, the one-loop quantum correction to this classical 
soliton mass is
\begin{eqnarray} 
M=\frac{1}{2}\int \frac{d\omega}{2\pi}\,{\rm tr}\,
\log\left[\omega^2-\frac{d^2}{dx^2}+V^{\prime\prime}(\phi_c(x))\right]
\label{oneloop}
\end{eqnarray} 
This expression must be suitably renormalized, but once this is done $M$ is  a pure number. Here $V^{\prime\prime}(\phi_c(x))$ means $\frac{d^2}{d\phi^2}V(\phi_c(x))$.
For the cases listed above, the bosonic fluctuation potential
$V^{\prime\prime}(\phi_c(x))$ is
\begin{eqnarray} 
V^{\prime\prime}(\phi_c(x))=\cases{1-\frac{3}{2}\,  {\rm
sech}^2(\frac{x}{2})\qquad   ;\phi^4 {\rm model}\cr 1-2\,{\rm
sech}^2(x)\qquad   ; {\rm SineGordon\,\, model}\cr
(1-\nu)\left({-1+2(1-\nu){\rm tanh}^2(x)+\nu\,{\rm tanh}^4(x)\over (1-\nu\, 
{\rm tanh}^2(x))^2}\right)
\qquad   ;  {\rm Jacobi \,\,model}}
\label{fl}
\end{eqnarray}
For the $\phi^4$ and Sine-Gordon cases, the fluctuation
potentials in  (\ref{fl}) are of the exactly solvable P\"oschl-Teller
form, which means that the quantum mass correction in (\ref{oneloop}) can
be computed exactly. The fluctuation potential for the Jacobi model is
shown in Figure 2; its spectral properties are not known exactly, and so we need an approximate method to compute the mass correction (\ref{oneloop}). Here we use the derivative expansion \cite{aitchison,chan}.

To implement the derivative expansion we first note that
$V^{\prime\prime}(\phi_c(x))\to 1$ as $x\to\pm\infty$, for the
fluctuation potentials appearing in (\ref{oneloop}). Moreover, this
background value of $1$ (equal to $m^2$ with mass scales reinstated) is
approached exponentially fast. So it is more natural to expand in terms
of the difference
\begin{eqnarray} 
W(x)=V^{\prime\prime}(\phi_c(x))- 1
\label{exp}
\end{eqnarray} 
and its $x$ derivatives. This leads to a generalized
derivative expansion \cite{chan} in which (\ref{oneloop}) is rewritten and expanded as
\begin{eqnarray} 
M[W]&=&-\int_{-\infty}^\infty\frac{d\omega}{2\pi}\,
\omega^2\, {\rm tr}\left(
\frac{1}{\omega^2+1-\frac{d^2}{dx^2}+W(x)}\right)\nonumber\\ &=&
\int_{-\infty}^\infty\frac{d\omega}{2\pi}\, \omega^2\, z^{\frac{3}{2}}\,
A(z,W]
\label{deriv}
\end{eqnarray} 
where $z\equiv 1/(\omega^2+1)$. In the derivative expansion
approach \cite{aitchison,chan}, the expression (\ref{deriv}) is
renormalized by dropping a divergent term that is independent of $W$ (this cancels in taking the difference with the $W=0$ case), and
a divergent term linear in $W$ (this corresponds to making the bosonic tadpole graph vanish, as in \cite{top,jaffe}). Then $A(z,W]$ is defined by the expansion 
\begin{eqnarray} 
A(z,W]=-\frac{1}{2}\,z\, \sum_{n=0}^\infty z^n\, a_n[W]
\label{func}
\end{eqnarray} 
where the expansion coefficients $a_n[W]$ are functionals of $W(x)$, the first
few of which are \cite{chan}:
\begin{eqnarray} a_0[W]&=&\frac{3}{8}\int_{-\infty}^\infty W^2\,
dx\nonumber\\ a_1[W]&=&-\frac{5}{32}\int_{-\infty}^\infty \left\{
2W^3+(W^\prime)^2\right\}\, dx\nonumber\\
a_2[W]&=&\frac{7}{128}\int_{-\infty}^\infty \left\{
5W^4+10W(W^\prime)^2+(W^{\prime\prime})^2\right\} \, dx\nonumber\\
a_3[W]&=&-\frac{9}{512}\int_{-\infty}^\infty \left\{14W^5+70
W^2(W^\prime)^2+14W(W^{\prime\prime})^2+(W^{\prime\prime\prime})^2\right\}
\, dx
\label{as}
\end{eqnarray}
Given the expansion (\ref{func}) for $A(z,W]$, the $\omega$ integrals in
(\ref{deriv}) can be performed to yield a simple formula for the soliton
mass correction:
\begin{eqnarray} 
M[W]=-\frac{1}{8\sqrt{\pi}}\,\sum_{n=0}^\infty
\frac{\Gamma(n+1)}{\Gamma(n+5/2)}\, a_n[W]
\label{formula}
\end{eqnarray} 
Thus, the derivative expansion computation of the soliton
mass correction reduces to the calculation of the $a_n[W]$ in (\ref{as})
for the given $W(x)$ in (\ref{exp}). We stress again that each of these
$a_n[W]$ is a pure number \cite{comment}. 

The functionals $a_n[W]$ can be obtained by a direct expansion of the resolvent
in (\ref{deriv}), as in \cite{chan}. However, it is easier
computationally to note that these functionals have simple recurrence
relations (these functionals also arise in asymptotic and WKB expansions
of discriminants \cite{braden,bd}, and in zeta function analysis
\cite{konoplich}). In fact, 
\begin{eqnarray} 
a_n[W]=(-1)^{n+1} (2n+3)\, \int_{-\infty}^\infty
r_{2n+4}(x)\, dx
\label{rs}
\end{eqnarray} where the functions $r_n(x)$ are defined by the following
recurrence relation:
\begin{eqnarray} 
r_0(x)&=&1\quad,\quad r_1(x)=0\quad,\quad
r_2(x)=-\frac{1}{2}W(x)\quad, \nonumber\\ r_n(x)&=&\frac{i}{2}\,
r_{n-1}^\prime(x)-\frac{1}{2} \sum_{j=1}^{n-1}r_j(x)\, r_{n-j}(x)\quad,\quad n\geq 3
\label{recurrence}
\end{eqnarray} 
This algorithm for generating the $r_{2n+4}(x)$ functions
is simple to implement using, for example, Mathematica or Maple.
Furthermore, all the integrals in (\ref{rs}) can be performed
analytically for each of the three soliton models listed in (\ref{pots}). 

So far, this algorithm could be applied to any suitable potential $W(x)$
in (\ref{deriv}), with $W(x)\to 0$ sufficiently quickly as
$x\to\pm\infty$. If the background is that of a soliton $\phi_c(x)$
satisfying (\ref{soliton}), so that $W(x)$ is given by (\ref{exp}), it is
more convenient to change variables in the integration in (\ref{rs}) from
$x$ to $\phi_c$, using (\ref{soliton}). For example:
\begin{eqnarray}
a_0[W]&=&\frac{3}{8}\int_{min}\frac{d\phi}{\sqrt{2V(\phi)}}\, W^2
\nonumber\\
a_1[W]&=&-\frac{5}{32}\int_{min}\frac{d\phi}{\sqrt{2V(\phi)}}\, \left\{
2W^3+2V(V^{(3)})^2\right\}\nonumber\\
a_2[W]&=&\frac{7}{128}\int_{min}\frac{d\phi}{\sqrt{2V(\phi)}}\, \left\{
5W^4+20WV(V^{(3)})^2+(V^{\prime}V^{(3)}+2 VV^{(4)})^2\right\}\nonumber\\
a_3[W]&=&-\frac{9}{512}\int_{min}\frac{d\phi}{\sqrt{2V(\phi)}}\,
\left\{14W^5+140W^2V(V^{(3)})^2+14W(V^\prime V^{(3)}+2V V^{(4)})^2\right. \nonumber\\
&&\hskip 3cm\left.+2V(V^{\prime\prime}V^{(3)} +3V^\prime V^{(4)}+2VV^{(5)})^2\right\} 
\label{revas}
\end{eqnarray} 
These modified expressions for $a_n[W]$ are straightforward to generate, either by converting the expressions in (\ref{as}), or by a new recurrence relation based on (\ref{recurrence}). In (\ref{revas}), both $V$ and $W=V^{\prime\prime}-1$ are treated as
functions of $\phi$, and the integration is between the neighboring
minima of $V(\phi)$ between which the soliton interpolates, just as in the standard calculation (\ref{cmasses}) of the classical mass. The advantages of this form (\ref{revas}) are : (i) the only input is the original potential
$V(\phi)$; (ii) the integrals are easier than those in (\ref{as}); (iii)
it organizes the expansion (\ref{formula}) as a derivative expansion, as
each $a_n$ in (\ref{revas}) involves only terms with $(2n+4)$ derivatives
of $V$ with respect to $\phi$. 

The derivative expansion calculation of the soliton mass correction
(\ref{formula}) can now be programmed as a simple algorithm: generate the expressions in (\ref{as}) or (\ref{revas}) using recursion relations, do the integrals, and then insert the $a_n$ into (\ref{formula}). We first test this algorithm on the well-known solvable cases. For the Sine-Gordon model one
finds $a_n^{\rm SG}=2$ for all $n$, confirming a result in  \cite{chan}.
Then, using our formula (\ref{formula}), we immediately obtain the
standard result \cite{rajaraman,rebbi}
\begin{eqnarray} 
M^{\rm SG}= -\frac{1}{4\sqrt{\pi}}\,\sum_{n=0}^\infty
\frac{\Gamma(n+1)}{\Gamma(n+5/2)}=-\frac{1}{\pi}
\label{sgmass}
\end{eqnarray} 
For the $\phi^4$ model, one finds $a_n^{\phi^4}=2+(1/4)^{n+1}$ for all $n$, again confirming a result in \cite{chan}. Then, from (\ref{formula}) we find the standard result
\cite{rajaraman,rebbi}
\begin{eqnarray} 
M^{\rm \phi^4}=-\frac{1}{\pi}
-\frac{1}{32\sqrt{\pi}}\,\sum_{n=0}^\infty
\frac{\Gamma(n+1)}{4^n\Gamma(n+5/2)}=\frac{1}{4\sqrt{3}}-\frac{3}{2\pi}
\label{ffmass}
\end{eqnarray}
One can speed up the convergence of the derivative expansion by separating out the contribution from the zero mode of the fluctuation potential \cite{chan}. This is achieved by writing $A(z,W]=(A(z,W]+\frac{1}{\omega^2})-\frac{1}{\omega^2}$ in (\ref{deriv}) and (\ref{func}), which separates out a $-\frac{1}{\pi}$ term, so that the mass correction (\ref{formula}) becomes
\begin{eqnarray} 
M[W]=-\frac{1}{\pi}-\frac{1}{8\sqrt{\pi}}\,\sum_{n=0}^\infty
\frac{\Gamma(n+1)}{\Gamma(n+5/2)}\, \left(a_n[W]-2\right)
\label{modformula}
\end{eqnarray}
In this form (\ref{modformula}), already the leading $(n=0)$ derivative expansion term gives the exact answer for the Sine-Gordon case. For the $\phi^4$ model, the leading term gives $10\%$ accuracy, the first correction gives better than $2\%$ accuracy, two corrections give $0.3\%$ accuracy, and three corrections give $0.05\%$ accuracy. This is impressive accuracy, especially considering $a_0, \dots, a_3$ can easily be calculated analytically.

For the Jacobi model there does not appear to be a simple closed formula for
the $a_n$. This is presumably a reflection of the fact that in this case,
unlike in the $\phi^4$ and Sine-Gordon cases,
$V^{\prime\prime}(\phi_c(x))$ is not an exactly solvable potential.
Nevertheless, we can compute various orders $a_n$ of the derivative expansion
expression with ease:

\begin{eqnarray} 
a_0^{\rm Lame}&=& - \frac{3}{128\nu^{3/2}}\left[2\sqrt{\nu}
(5 - 38\nu + 5\nu^2) -(5 + 3\nu + 3\nu^2 + 5\nu^3) 
\log\left(\frac{1 + \sqrt{\nu}}{1-\sqrt{\nu}}\right)\right] 
\nonumber\\ 
a_1^{\rm Lame}&=& \frac{1}{12288\nu^{5/2}}\left[-2\sqrt{\nu}(1785 + 520\nu -11746\nu^2 + 520\nu^3 + 1785\nu^4) \right. \nonumber\\
&&\hskip 2cm\left.+ 15(119 - 5\nu + 14\nu^2 + 14\nu^3 - 5\nu^4 + 119\nu^5)\log\left(\frac{1+\sqrt{\nu}}{1-\sqrt{\nu}}\right)\right] \nonumber\\ 
a_2^{\rm Lame}&=& \frac{1}{3932160 \nu^{7/2}} 
\left[-2\sqrt{\nu}(1323945 - 683970\nu + 146839\nu^2 -3800572\nu^3 + 
       146839\nu^4 - 683970\nu^5 + 1323945\nu^6) \right.\nonumber\\
&&\left. + 105(12609 - 10717\nu + 2449\nu^2 + 779\nu^3 + 779\nu^4 + 
       2449\nu^5 - 10717\nu^6 + 12609\nu^7) \log\left(\frac{1+\sqrt{\nu}}{1-\sqrt{\nu}}\right)\right] \nonumber\\ 
a_3^{\rm Lame}&=& -\frac{3}{587202560\nu^{9/2}}\left[2\sqrt{\nu}
        (258321525 - 355550860\nu + 143327940\nu^2 - 681332\nu^3 \right.\nonumber\\
&&\hskip 2cm\left. - 194557554\nu^4 - 681332\nu^5 + 143327940\nu^6 - 355550860\nu^7 + 258321525\nu^8)\right.\nonumber\\
&&\hskip 2cm\left. - 105(2460205 - 4206267\nu + 2275076\nu^2 - 275052\nu^3 - 24586\nu^4 - 24586\nu^5\right.\nonumber\\
&&\hskip 2cm \left. - 275052\nu^6 + 2275076\nu^7 - 4206267\nu^8 + 2460205\nu^9) \log\left(\frac{1+\sqrt{\nu}}{1-\sqrt{\nu}}\right)\right]
\label{lameas}
\end{eqnarray}
Notice the appearance of the classical soliton mass
$\frac{4}{\sqrt{\nu}}\log\left(\frac{1+\sqrt{\nu}}{1-\sqrt{\nu}}\right)$
in these expressions for the one-loop quantum correction. 

Figure 3 shows the successive derivative expansion approximations to the
quantum mass correction, as a function of the elliptic parameter $\nu$.
The expansion appears to be convergent for each $\nu$. Furthermore, a nontrivial test of the results in (\ref{lameas}) is that they must reduce to the Sine-Gordon results as $\nu\to 0$. Indeed, from (\ref{lameas}) we verify that for each $n$
\begin{eqnarray} a_n^{\rm Lame}\to 2 \quad {\rm as}\quad \nu\to 0
\label{red}
\end{eqnarray}  
On the other hand, as $\nu\to 1$ the quantum mass correction diverges, just
as does the classical Jacobi soliton mass in (\ref{cmasses}). This singular behavior is due to the fact that as $\nu\to 1$ the period of the Jacobi potential diverges logarithmically, so neighboring vacua become infinitely separated, and the asymptotic limits of the classical soliton $\phi_c(x)$ in (\ref{sols}) diverge.

We now consider the $N=1$ supersymmetric extension of the bosonic soliton
model (\ref{blag}). Consider
\begin{eqnarray} 
{\cal L}_{\rm SUSY}=\frac{1}{2}(\partial_\mu\phi)
(\partial^\mu\phi)-\frac{1}{2}U^2(\phi)
+\frac{i}{2}\bar{\psi}\partial \hskip-6pt /\psi 
-\frac{1}{2}U^\prime(\phi) \bar{\psi}\psi
\label{susylag}
\end{eqnarray} 
where $\psi$ is a Majorana fermion and
\begin{eqnarray} 
U(\phi)=\cases{\frac{1}{2}\left(\phi^2-1\right)\qquad ; \phi^4 {\rm model}\cr 
-2\, {\rm sin}(\frac{\phi}{2})\qquad   ; {\rm SineGordon\,\, model}\cr 
-2\,{\rm sn}\left(\frac{\phi}{2}|\nu\right)\qquad  ;  {\rm Jacobi \,\,model}}
\label{upots}
\end{eqnarray} 
The bosonic potential is $V(\phi)=\frac{1}{2}U^2(\phi)$, solitons are given by $\frac{d}{dx}\phi_c=-U(\phi_c(x))$, and the boson-fermion coupling is 
\begin{eqnarray} 
V_F(\phi)=U^\prime(\phi)
\label{fpot}
\end{eqnarray}
 
\begin{figure}[ht]
\vskip -1cm
\centering{\epsfig{file=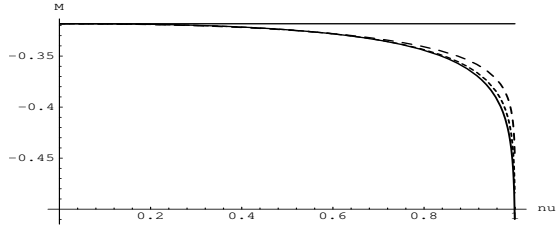,width=5in,height=4.5in}}
\vskip -2.5in
\caption{The successive derivative expansion approximations for the quantum mass correction in (\protect{\ref{modformula}}) for the Jacobi model, using the coefficients $a_n$ in (\protect{\ref{lameas}}), plotted as a function of the elliptic modular parameter $\nu$. The dashed curve is the leading $(n=0)$ contribution, the dotted curve includes the first two terms, while the solid curve includes the first three terms, and is indistinguishable on this scale from that with the first four terms. The solid horizontal line is the Sine-Gordon mass correction $M^{SG}=-\frac{1}{\pi}$, to which the Jacobi model reduces when $\nu=0$.}
\end{figure}

The fermionic contribution to the one-loop soliton mass
correction is
\begin{eqnarray} 
M_F&=&i\,{\rm tr}\,\log\left[i\partial \hskip-6pt
/ -V_F(\phi_c(x))\right]\nonumber\\ 
&=&-\frac{1}{2}\left\{M[W_+]+M[W_-]\right\}
\label{fmass}
\end{eqnarray} 
where $M[W]$ is the bosonic mass correction in (\ref{deriv}) and
\begin{eqnarray} 
W_\pm(x)\equiv \left[V_F(\phi_c(x))\right]^2\pm\frac{d}{dx}V_F(\phi_c(x)) -1
\label{susyq}
\end{eqnarray} 
But with the SUSY coupling (\ref{fpot}), $W_-(x)$ is equal to the bosonic potential $W(x)$:
\begin{eqnarray} 
W_-(x)=W(x)\equiv \frac{d^2}{d\phi^2}V(\phi_c(x))-1
\label{eq}
\end{eqnarray} 
Thus the {\it total} one-loop mass correction, including both
bosonic and fermionic contributions, is 
\begin{eqnarray} 
M_{\rm total}=\frac{1}{2}\left\{M[W_-]-M[W_+]\right\}
\label{diff}
\end{eqnarray} 
That is, the net mass correction is (half) the difference
of two bosonic mass corrections, for $W_\mp(x)$ respectively. 

Now notice the remarkable fact that if we compute this difference using
the derivative expansion (\ref{formula}), then term by term each of the
functionals $a_n[W_-]-a_n[W_+]$ obtained from (\ref{as}) is an integral
of a total derivative. For example,
\begin{eqnarray} 
a_0[W_-]-a_0[W_+]&=&\frac{3}{8}\int_{-\infty}^\infty
\frac{4}{3}\frac{d}{dx} \left\{3V_F-V_F^3\right\}\, dx\nonumber\\
a_1[W_-]-a_1[W_+]&=&-\frac{5}{32}\int_{-\infty}^\infty
\frac{12}{5}\frac{d}{dx} \left\{ -5V_F+ \frac{10}{3}V_F^3-V_F^5-\frac{5}{3}V_F(V_F^\prime)^2\right\}\, dx \nonumber\\
a_2[W_-]-a_2[W_+]&=&\frac{7}{128}\int_{-\infty}^\infty
\frac{40}{7}\frac{d}{dx} \left\{ 7V_F-7V_F^3+\frac{21}{5}V_F^5-V_F^7+7(V_F-V_F^3)(V_F^\prime)^2-
\frac{7}{5}V_F^{\prime\prime} (V_F^\prime)^2 -\frac{7}{10}V_F(V_F^{\prime\prime})^2\right\} \, dx\nonumber\\
a_3[W_-]-a_3[W_+]&=&-\frac{9}{512}\int_{-\infty}^\infty
\frac{140}{9}\frac{d}{dx} \left\{ -9V_F+12V_F^3-\frac{54}{5}V_F^5+\frac{36}{7}V_F^7-V_F^9 -18V_F(1-V_F^2)^2(V_F^\prime)^2 -9V_F(V_F^\prime)^4\right.\nonumber\\
&&\left. +\frac{36}{5}(1-V_F^2)(V_F^\prime)^2V_F^{\prime\prime} +\frac{18}{5}V_F(1-V_F^2)(V_F^{\prime\prime})^2+
\frac{18}{35}(V_F^{\prime\prime})^3 -\frac{54}{35}V_F^\prime V_F^{\prime\prime} V_F^{\prime\prime\prime} -\frac{9}{35}V_F(V_F^{\prime\prime\prime})^2\right\}
\, dx\nonumber\\
\label{diffas}
\end{eqnarray} 
In (\ref{diffas}), $V_F^\prime$ means
$\frac{d}{dx}V_F(\phi_c(x))$. It is straightforward to generate these expressions to a desired order using Mathematica or Maple.  

Due to this dramatic simplification, the differences $a_n[W_-]-a_n[W_+]$
are determined solely by the asymptotic values (at $x=\pm\infty$) of
$V_F(x)$ and its $x$ derivatives. But for a solitonic background, $V_F(\phi_c(x))$ approaches $\pm 1$ exponentially fast, so 
\begin{eqnarray} 
V_F(\phi_c(x))&=&\pm 1 \quad {\rm at} \quad x=\pm\infty\nonumber\\ 
V_F^{(n)}(\phi_c(x))&=&0 \quad {\rm at} \quad
x=\pm\infty\quad, \quad n\geq 1
\label{asymptotic}
\end{eqnarray} 
For example, these conditions are clearly satisfied for the $\phi^4$,
Sine-Gordon and Jacobi models, 
\begin{eqnarray} 
V_F(\phi_c(x))=\cases{{\rm tanh}(\frac{x}{2})\qquad   ;\phi^4 {\rm  model}\cr 
{\rm tanh}(x)\qquad   ; {\rm SineGordon\,\,model}\cr
\frac{(1-\nu)\, {\rm tanh}(x)}{1-\nu\,{\rm tanh}^2(x)}
\qquad   ;  {\rm Jacobi \,\,model}}
\label{fpotegs}
\end{eqnarray}
Inserting the asymptotic conditions (\ref{asymptotic}) into (\ref{diffas}), the derivatives of $V_F$ do not contribute, and we discover that {\it the differences are independent of n}: 
\begin{eqnarray}
a_n[W_-]-a_n[W_+]=2\quad, \quad {\rm for\,\, all\,\, n}
\label{ans}
\end{eqnarray} 
Thus, combining the derivative expansion formula
(\ref{formula}) with the expression (\ref{diff}) for the total mass
correction, we find (in units of $m$)
\begin{eqnarray} M_{\rm total}=-\frac{1}{8\sqrt{\pi}}\,\sum_{n=0}^\infty
\frac{\Gamma(n+1)}{\Gamma(n+5/2)}=-\frac{1}{2\pi}
\label{totalmass}
\end{eqnarray} This is in agreement with recent analyses \cite{top,jaffe}
using topological boundary conditions and phase shift methods. We note
that in the derivative expansion approach, this result is quite general,
relying only on the asymptotic values of $V_F(\phi_c(x))$ and its $x$
derivatives.

\vskip 1cm

This work has been supported in part by the U.S. Department of Energy  
grant DE-FG02-92ER40716.00.

\end{document}